\documentclass[aps,rpl,preprint,groupedaddress]{revtex4}
\begin{document}

\title{Hyperfine interaction induced decoherence and deterministic teleportation of electrons in a quantum dot nanostructure}

\author{Siqing Yu}
\affiliation{Hwa Chong Institution, 661 Bukit Timah Road, Singapore 269734, Singapore}

\author{Yechao Zhu}
\affiliation{Hwa Chong Institution, 661 Bukit Timah Road, Singapore 269734, Singapore}

\author{Ye Yeo}
\affiliation{Department of Physics, National University of Singapore, 10 Kent Ridge Crescent, Singapore 119260, Singapore}

\begin{abstract}
Recently, de Visser and Blaauboer [Phys. Rev. Lett. {\bf 96}, 246801 (2006)] proposed the most efficient deterministic teleportation protocol $\cal T$ 
for electron spins in a semiconductor nanostructure consisting of a single and a double quantum dot.  However, it is as yet unknown if $\cal T$ can be 
completed before decoherence sets in.  In this paper we analyze the detrimental effect of nuclear spin baths, the main source of decoherence, on 
$\cal T$.  We show that nonclassical teleportation fidelity can be achieved with $\cal T$ provided certain conditions are met.  Our study indicates that realization of 
quantum computation with quantum dots is indeed promising.
\end{abstract}

\maketitle

Quantum teleportation \cite{Bennett} provides a means to the complete transfer of quantum information from one particle to another.  It is a possible
primitive for large-scale quantum computers \cite{Gottesman}.  Recently, de Visser and Blaauboer \cite{Visser} proposed the most efficient 
deterministic teleportation protocol $\cal T$ for electron spins in a semiconductor nanostructure consisting of a single and a double quantum dot.  
This is an important step in the realization of quantum computation \cite{Nielsen} with quantum dots, first proposed by Loss and DiVincenzo 
\cite{DiVincenzo}.

The total duration of the teleportation process is estimated to be between 150 and 250 ns \cite{Visser}.  This estimate is based on the experimental 
fact that the duration of an electron-spin resonance (ESR) rotation is roughly 50 ns for magnetic fields of 1 mT.  For typical GaAs quantum dots, due to 
interaction with the environment, the electron spin decoheres very quickly, on a time scale of order of nanoseconds \cite{Johnson, Koppens, Petta}.  It 
is as yet unknown if the protocol $\cal T$ can be completed before decoherence sets in.  In the experimentally relevant regime of moderate magnetic 
fields (less than a few T) and temperatures of tens or hundreds of mK, the decoherence is largely due to the bath of $N$ nuclear spins.  For 
temperatures as low as 100 mK, in the case of GaAs quantum dots, the state of the spin bath is given by 
$\rho_{\rm bath} = ({\bf 1}_1 \otimes \cdots \otimes {\bf 1}_N)/2^N$ \cite{Koppens, Petta}.  Here, ${\bf 1}_k$ is the identity matrix for the {\it k}th 
nuclear spin.

In this paper, we show that one is able to achieve nonclassical teleportation fidelity with $\cal T$ in the presence of decoherence caused by 
the nuclear spin baths.  For instance, provided we have the means to create an ESR square pulse of 6 mT amplitude, we could complete $\cal T$ in 
$\tau_{\rm tot} \approx 7.3\ {\rm ns}$ and attain a teleportation fidelity of $\Phi \approx 93\%$.  To this end, we consider each electron coupled via 
the contact hyperfine interaction to the bath of $N$ nuclear spins and to an external magnetic field $B_0$.  The Hamiltonian is given by
\begin{equation}\label{Hamiltonian}
{\cal H} = H_0S_{\alpha} + A\vec{S}\cdot\vec{I}.
\end{equation}
$H_0 = 2\mu_bB_0$, where the Bohr magneton $\mu_b = e\hbar/2m_e \approx 5.788 \times 10^{-5} {\rm eV}/{\rm T}$.  $\vec{S}$ is the operator of the 
electron spin and $\alpha$ could be $x$, $y$, or $z$, depending on if $B_0$ is applied along the $x$-, $y$-, or $z$-direction.  With $\vec{I}_k$ 
the operator of the $k$th bath spin ($k = 1, 2, \cdots, N$), the total nuclear spin of the bath $\vec{I} = \sum_k\vec{I}_k$.  $A$ is the contact 
hyperfine coupling.  For a typical GaAs quantum dot with the electron delocalized over $N = 10^6$ nuclear spins, $A \approx 10^{-10}$ eV \cite{Wenxian}.  
We note that the accuracy of our prediction of $\tau_{\rm tot}$ depends on that of $A$ and therefore $N$.  Hence, precise measurements of 
$\tau_{\rm tot}$ and $\Phi$ may yield more accurate values of $A$ and $N$.  In the following, we present our results alongside the steps in $\cal T$.

{\it Step 0.}  At time $t = 0$, we suppose Alice and Bob share a pair of qubits, $A_2$ and $B$, in a maximally entangled Bell singlet state 
$|\Psi\rangle_{A_2B} = (|01\rangle_{A_2B} - |10\rangle_{A_2B})/\sqrt{2}$.  And, the qubit $A_1$ that Alice is to teleport to Bob is in some arbitrary 
pure state $|\psi\rangle_{A_1} = (a|0\rangle_{A_1} + b|1\rangle_{A_1})$, where $a = \cos\theta/2$ and $b = \exp i\phi\sin\theta/2$ with 
$0 \leq \theta \leq \pi$ and $0 \leq \phi \leq 2\pi$.  The initial total state of qubits $A_1$, $A_2$ and $B$ is thus given by
\begin{equation}
|\Gamma_0\rangle_{A_1A_2B} \equiv |\psi\rangle_{A_1} \otimes |\Psi\rangle_{A_2B}.
\end{equation}

{\it Step 1.}  For electron spins in quantum dots, since there is no measurement technique available for full Bell measurements, the first step in 
$\cal T$ unitarily transforms $|\Gamma_0\rangle_{A_1A_2B}$ into \cite{Visser}
\begin{equation}
|\Gamma_1\rangle_{A_1A_2B} \equiv ({\cal S}_{A_1A_2} \otimes {\cal I}_B)(|\psi\rangle_{A_1} \otimes |\Psi\rangle_{A_2B}),
\end{equation}
where the $\sqrt{{\rm SWAP}}$ operator
\begin{equation}
{\cal S} = \frac{e^{\frac{i\pi}{8}}}{\sqrt{2}}\left(\begin{array}{cccc}
1- i &  0 &  0 & 0 \\
0    &  1 & -i & 0 \\
0    & -i &  1 & 0 \\
0    &  0 &  0 & 1 - i
\end{array}\right).
\end{equation}
This is achieved by turning on the exchange interaction between electron spins in quantum dots, described by the Heisenberg Hamiltonian 
${\cal H}_{\rm ex} = J\vec{S}_{A_1}\cdot\vec{S}_{A_2}$, where $J$ is the time-dependent exchange energy.  Clearly, ${\cal H}_{\rm ex}$ does not commute 
with $A\vec{S}\cdot\vec{I}$ in Eq.({\ref{Hamiltonian}).  Mathematically, it is therefore difficult to consider the effect of the nuclear spin baths on 
the dynamics of qubits $A_1$ and $A_2$.  However, we note that the duration of the $\sqrt{\rm SWAP}$ operation is only about 0.18 ns \cite{Petta}.  
This is tiny compared to 10 ns \cite{Koppens, Petta}, the time interval necessary for the baths to have significant influence.  So, we ignore the baths 
and have $|\Gamma_1\rangle_{A_1A_2B} = e^{\frac{i\pi}{8}}/2[(1 - i)a|001\rangle_{A_1A_2B} - a|010\rangle_{A_1A_2B} - ib|011\rangle_{A_1A_2B} 
+ ia|100\rangle_{A_1A_2B} + b|101\rangle_{A_1A_2B} - (1 - i)b|110\rangle_{A_1A_2B}]$.

{\it Step 2.}  Immediately after the $\sqrt{\rm SWAP}$ operation, Alice subjects her qubit $A_1$ to a single-qubit rotation by $\vartheta$ ($= -\pi$) 
about an axis $\hat{n} = (\cos\varphi, \sin\varphi, 0)$:
\begin{equation}
{\cal R}^{(\hat{n})}(\vartheta) = \left(\begin{array}{cc}
\cos\frac{\vartheta}{2}               & -ie^{-i\varphi}\sin\frac{\vartheta}{2} \\
-ie^{i\varphi}\sin\frac{\vartheta}{2} & \cos\frac{\vartheta}{2}
\end{array}\right).
\end{equation}
Ideally, without considering the effect of the nucelar spin baths, we would have
\begin{equation}
|\Gamma_2\rangle_{A_1A_2B} \equiv [{\cal R}^{\hat{n}}_{A_1}(-\pi) \otimes {\cal I}_{A_2B}]|\Gamma_1\rangle_{A_1A_2B}.
\end{equation}
Here, for simplicity, we consider single-qubit rotation about the $x$-axis or $\varphi = 0$.  Consequently, we have
$|\Gamma_2\rangle_{A_1A_2B} = i\exp(i\pi/8)[ia|000\rangle_{A_1A_2B} + b|001\rangle_{A_1A_2B} - (1 - i)b|010\rangle_{A_1A_2B}
+ (1 - i)a|101\rangle_{A_1A_2B} - a|110\rangle_{A_1A_2B} - ib|111\rangle_{A_1A_2B}]/2$.  This is effected by applying an ESR pulse to qubit $A_1$, 
described by the Hamiltonian ${\cal H}_{\rm rot} = H_0S_x$.  Since the duration of an ESR rotation is about 50 ns for magnetic 
fields of 1 mT, it is important we take into account the decohering effects of the individual nuclear spin baths.  For qubit $A_1$, this is 
accomplished by considering the unitary dynamics of $A_1$ and its associated spin bath, which is generated by ${\cal H}$ in Eq.(\ref{Hamiltonian}) with
$\alpha = x$.  The open system dynamics of $A_1$ can then be determined by tracing over the spin bath degrees of freedom.  This is described by a 
completely positive, trace-preserving map ${\cal E}^x$, defined by
\begin{eqnarray}\label{x}
|0\rangle\langle 0| & \longrightarrow & \frac{2 + R_1}{4}|0\rangle\langle 0| - i\frac{R_2}{4}    |0\rangle\langle 1| 
                                     + i\frac{R_2}{4}    |1\rangle\langle 0| +  \frac{2 - R_1}{4}|1\rangle\langle 1|, \nonumber \\
|0\rangle\langle 1| & \longrightarrow & -i\frac{R_2}{4}         |0\rangle\langle 0| +  \frac{2 + R_1 - 4W}{4}|0\rangle\langle 1|
                                        + \frac{2 - R_1 - 4W}{4}|1\rangle\langle 0| + i\frac{R_2}{4}         |1\rangle\langle 1|, \nonumber \\
|1\rangle\langle 0| & \longrightarrow & i\frac{R_2}{4}          |0\rangle\langle 0| +  \frac{2 - R_1 - 4W}{4}|0\rangle\langle 1|
                                        + \frac{2 + R_1 - 4W}{4}|1\rangle\langle 0| - i\frac{R_2}{4}         |1\rangle\langle 1|, \nonumber \\
|1\rangle\langle 1| & \longrightarrow & \frac{2 - R_1}{4}|0\rangle\langle 0| + i\frac{R_2}{4}    |0\rangle\langle 1| 
                                     - i\frac{R_2}{4}    |1\rangle\langle 0| +  \frac{2 + R_1}{4}|1\rangle\langle 1|.
\end{eqnarray}
Here,
\begin{eqnarray}\label{alpha}
R_1 & \approx & 2W + 2\exp\left(-\frac{Nt^2}{8}\right)\left[\cos(\lambda t) - \frac{N t}{4 \lambda}\sin(\lambda t)\right], \nonumber \\
R_2 & \approx & 2\exp\left(-\frac{Nt^2}{8}\right)\left[\left(\frac{N}{4\lambda^2} - 1\right)\sin(\lambda t) - \frac{N t}{4 \lambda}\cos(\lambda t)\right],
\end{eqnarray}
and with $\rm erf$ being the error function \cite{Wenxian},
\begin{eqnarray}\label{W}
W & \approx & \frac{N}{4\lambda^2}\left[1 - \exp\left(-\frac{Nt^2}{8}\right)\cos(\lambda t)\right] + i\sqrt{\frac{\pi N^3}{512\lambda^6}}
\exp\left(-\frac{2\lambda^2}{N}\right) \nonumber \\
& & \times \left[{\rm erf}\left(\frac{Nt - i4\lambda}{2\sqrt{2N}}\right) 
- {\rm erf}\left(\frac{Nt + i4\lambda}{2\sqrt{2N}}\right) + 2{\rm erf}\left(\frac{i4\lambda}{2\sqrt{2N}}\right)\right].
\end{eqnarray}
To analyze the effect of the nuclear spin bath on the single-qubit rotation, we define the {\em gate fidelity} averaged 
over all possible $|\psi\rangle$:
\begin{eqnarray}\label{sigmaxfidelity}
{\cal G}_x(\pi) & \equiv & \max_{t = \tau_x(\pi)}\left\{
\frac{1}{4\pi}\int^{2\pi}_0\int^{\pi}_0\langle\chi|{\cal E}^x(|\psi\rangle\langle\psi|)|\chi\rangle\sin\theta d\theta d\phi\right\} \nonumber \\
& = & \max_{t = \tau_x(\pi)}\left\{\frac{1}{6}(4 - R_1 - 2W)\right\},
\end{eqnarray}
where $|\chi\rangle = {\cal R}^{(\hat{x})}(\pi)|\psi\rangle$, and
\begin{equation}
{\cal E}^{\alpha}(|\psi\rangle\langle\psi|) = {\rm tr}_{\rm bath}\left[\exp\left(-\frac{it}{\hbar}{\cal H}\right)
(|\psi\rangle\langle\psi| \otimes \rho_{\rm bath})\exp\left(\frac{it}{\hbar}{\cal H}\right)\right].
\end{equation}
In Eqs.(\ref{alpha}) and (\ref{W}), we have defined $\lambda \equiv H_0/A$ and from there on we set both $A = 1$ and $\hbar = 1$.  From Table I, we 
note that with smaller $B_0$, not surprisingly, it takes a longer time to reach a lower gate fidelity.  In particular, it takes about 13 ns to reach a 
gate fidelity of 0.72.  We emphasize that this calculation is based on the assumptions that $N = 10^6$, $A \approx 10^{-10}$ eV, and $B_0$ is in the 
form of a square pulse.  These should account for how it differs from the experimentally measured 50 ns duration for an ESR rotation.  While qubit 
$A_1$ is being rotated, the dynamics of qubits $A_2$ and $B$ are also affected by their individual nuclear spin baths.  The system with bath dynamics is 
generated by $\cal H$ with $B_0 = 0$.  The open system dynamics of $A_2$ or $B$ can again be derived by tracing over the respective bath degrees of 
freedom, and is given by another completely positive, trace-preserving map ${\cal E}^z$, defined by
\begin{eqnarray}\label{nofield}
|0\rangle\langle 0| & \longrightarrow & (1 - Z) |0\rangle\langle 0| + Z|1\rangle\langle 1|, \nonumber \\
|0\rangle\langle 1| & \longrightarrow & \gamma  |0\rangle\langle 1|, \nonumber \\
|1\rangle\langle 0| & \longrightarrow & \gamma^*|1\rangle\langle 0|, \nonumber \\
|1\rangle\langle 1| & \longrightarrow & Z |0\rangle\langle 0| + (1 - Z)|1\rangle\langle 1|.
\end{eqnarray}
$\gamma \equiv 1/2\lim_{B_0 \rightarrow 0}R_1$ and $Z \equiv \lim_{B_0 \rightarrow 0}W$.  The conjugate of $\gamma$, $\gamma^* = \gamma$ since $\gamma$ 
is real.  To determine the effect of the spin baths on the qubits, we calculate
\begin{equation}
\frac{1}{4\pi}\int^{2\pi}_0\int^{\pi}_0\langle\psi|{\cal E}^z(|\psi\rangle\langle\psi|)|\psi\rangle\sin\theta d\theta d\phi =
\frac{1}{3}(2 + \gamma - Z).
\end{equation}
For $t = 0.18$ ns, $(2 + \gamma - Z)/3 \approx 0.999907$, which justifies the above assumption that the spin baths have negligible effects during the 
$\sqrt{\rm SWAP}$ operation.  This does not apply during the single-qubit rotation, especially when $B_0$ is small.  Therefore, we have a mixed state 
given by the density operator $\Gamma_2$ instead of $|\Gamma_2\rangle_{A_1A_2B}$.  $\Gamma_2$ is obtained by applying ${\cal E}^x$ and ${\cal E}^z$'s 
to $|\Gamma_1\rangle_{A_1A_2B}\langle\Gamma_1|$:
\begin{equation}
\Gamma_2 = ({\cal E}^x_{A_1} \otimes {\cal E}^z_{A_2} \otimes {\cal E}^z_B)(|\Gamma_1\rangle_{A_1A_2B}\langle\Gamma_1|).
\end{equation}

\begin{table}\label{I}
\begin{tabular}{|c|c|c|c|c|c|c|} \hline
$B_0$/mT                         & 1    & 2    & 3    & 4    & 5    & 6    \\ \hline
${\cal G}_x(\pi)$                & 0.72 & 0.89 & 0.94 & 0.97 & 0.98 & 0.98 \\ \hline
Duration, $\tau_x(\pi)$/ns       & 13   & 8.2  & 5.7  & 4.4  & 3.5  & 2.9  \\ \hline
\end{tabular}  \\
Table I: \quad Gate fidelities for different $B_0$'s. \par
\end{table}

{\it Step 3.}  Alice applies another $\sqrt{\rm{SWAP}}$ operation to qubits $A_1$ and $A_2$, and completes the unitary transformation necessary to 
turn the Bell basis into the standard basis $\{|00\rangle, |01\rangle, |10\rangle, |11\rangle\}$ (see Ref.\cite{Visser}).  Again, ideally we would have
\begin{eqnarray}\label{Gamma30}
|\Gamma_3\rangle_{A_1A_2B} & \equiv & ({\cal S}_{A_1A_2} \otimes {\cal I}_B)|\Gamma_2\rangle_{A_1A_2B} \nonumber \\
& = & \frac{i}{2}\sum^1_{j, k = 0}|jk\rangle_{A_1A_2} \otimes U^{\dagger}_{jk}|\psi\rangle_B.
\end{eqnarray}
The unitary transformations $U_{jk}$ are defined in Table II.  However, taking into account the effects of the nuclear spin baths, we instead have
\begin{equation}\label{Gamma31}
\Gamma_3 = ({\cal S}_{A_1A_2} \otimes {\cal I}_B)\Gamma_2({\cal S}^{\dagger}_{A_1A_2} \otimes {\cal I}_B)
\end{equation}
As in Step 1, we assume that the nuclear spin baths to have negligible effects here.

\begin{table}\label{II}
\begin{tabular}{|c|c|} \hline
Alice's measurement result $jk$ & Bob's unitary recovery operation $U_{jk}$ \\ \hline
00                & $U_{00} = e^{-i\pi/4}{\cal R}^{\hat{x}}(\frac{\pi}{2}){\cal R}^{\hat{y}}(\frac{\pi}{2}){\cal R}^{\hat{x}}(-\frac{\pi}{2})$ \\ \hline
01                & $U_{01} = -e^{i\pi/4}{\cal R}^{\hat{x}}(\frac{\pi}{2}){\cal R}^{\hat{y}}(\frac{\pi}{2}){\cal R}^{\hat{x}}(\frac{\pi}{2})$ \\ \hline
10                & $U_{10} = e^{i\pi/4}{\cal R}^{\hat{x}}(\frac{\pi}{2}){\cal R}^{\hat{y}}(-\frac{\pi}{2}){\cal R}^{\hat{x}}(\frac{\pi}{2})$ \\ \hline
11                & $U_{11} = -e^{-i\pi/4}{\cal R}^{\hat{x}}(-\frac{\pi}{2}){\cal R}^{\hat{y}}(\frac{\pi}{2}){\cal R}^{\hat{x}}(\frac{\pi}{2})$ \\ \hline
\end{tabular} \\
Table II: \quad Alice's measurement outcomes and Bob's corresponding unitary recovery operations. \par
\end{table}

{\it Step 4.} Alice performs single-qubit projective measurements on her qubits, $A_1$ and $A_2$, in the standard basis $\{|0\rangle, |1\rangle\}$.  
This is described by the following set of projectors.
\begin{equation}
\{M_{00} = |00\rangle_{A_1A_2}\langle00|, M_{01} = |01\rangle_{A_1A_2}\langle01|, 
  M_{10} = |10\rangle_{A_1A_2}\langle10|, M_{11} = |11\rangle_{A_1A_2}\langle11|\}
\end{equation}
From Eq.(\ref{Gamma30}), we can easily deduce the resulting states of Bob's qubit corresponding to each of Alice's measurement outcomes.  Similarly, we
can derive from Eq.(\ref{Gamma31}) the state of qubit $B$ given that Alice obtains the measurement outcome $jk$:
\begin{equation}
\sigma^{jk}_B = \frac{1}{p_{jk}}{\rm tr}_{A_1A_2}(M_{jk}\Gamma_3).
\end{equation}
$j, k \in \{0, 1\}$ and $p_{jk} = {\rm tr}[M_{jk}\Gamma_3]$ is the probability of outcome $jk$.  Assuming single-shot readout is fast, we can neglect 
the spin baths here too.  The fidelity of $\sigma^{jk}$ with respect to $U^{\dagger}_{jk}|\psi\rangle$ averaged over all possible $jk$'s and 
$|\psi\rangle$ is given by
\begin{eqnarray}
{\cal F} & \equiv & \max_{t = \tau}\left\{\frac{1}{4\pi}\int^{2\pi}_0\int^{\pi}_0
\sum^1_{i, j = 0}p_{ij}\langle\psi|U_{ij}\sigma^{ij}U^{\dagger}_{ij}|\psi\rangle\sin\theta d\theta d\phi\right\} \nonumber \\
& = & \max_{t = \tau}\left\{\frac{1}{2} + \frac{\gamma}{12}[1 - 2Z + \gamma(1 - 2W)] + \frac{R_1}{24}[8Z(1 - Z) - \gamma(1 + \gamma) - 2]\right\}.
\end{eqnarray}
Comparing Table III with Table I, we note that the single-qubit projective measurements clearly ``amplify" the noise and thus we have a fidelity smaller 
than the corresponding gate fidelity ${\cal G}_x(\pi)$.  For instance, when $B_0 = 6\ {\rm mT}$, it takes about 2.9 ns to reach a fidelity of 0.94.  
This effect is especially serious when $B_0$ is small - it takes about 9.4 ns to reach a fidelity of only 0.57 if $B_0 = 1\ {\rm mT}$.  We also observe 
interestingly that when $B_0$ is small, i.e., when $\tau_x(\pi)$ is close to or greater than 10 ns, we have $\tau < \tau_x(\pi)$.  It means that one is 
able to achieve a higher $\cal F$ before time $t = \tau_x(\pi)$, when the decohering effects of the baths become substantial.

\begin{table}\label{III}
\begin{tabular}{|c|c|c|c|c|c|c|} \hline
$B_0$/mT     & 1    & 2    & 3    & 4    & 5    & 6    \\ \hline
$\cal F$     & 0.57 & 0.72 & 0.83 & 0.89 & 0.92 & 0.94 \\ \hline
Duration, $\tau$/ns  & 9.4  & 6.8  & 5.2  & 4.1  & 3.4  & 2.9  \\ \hline
\end{tabular} \\
Table III: \quad Fidelities of $\sigma^{jk}$ with respect to $U^{\dagger}_{jk}|\psi\rangle$ averaged over all possible $jk$'s and 
$|\psi\rangle$. \par
\end{table}

{\it Step 5.}  It is clear from Eq.(\ref{Gamma30}) that, in the ideal case, upon receiving Alice's measurement result, Bob can always recover 
$|\psi\rangle_B$ by applying to his qubit $B$ appropriate unitary operation $U_{jk}$ given in Table II.  Each $U_{jk}$ is composed of single-qubit 
rotations ${\cal R}^{(\hat{x})}$ and ${\cal R}^{(\hat{y})}$ about the $x$- and $y$-axes respectively.  As in Step 2, ${\cal R}^{(\hat{y})}$ is generated 
by the Hamiltonian ${\cal H}_{\rm rot} = H_0S_y$.  It is again important to consider the nuclear spin bath.  The open system dynamics of $B$, in this case, is described by 
the completely positive, trace-preserving map ${\cal E}^y$, defined by
\begin{eqnarray}
|0\rangle\langle 0| & \longrightarrow & \frac{2 + R_1}{4}|0\rangle\langle 0| - \frac{R_2}{4}    |0\rangle\langle 1| 
                                      - \frac{R_2}{4}    |1\rangle\langle 0| + \frac{2 - R_1}{4}|1\rangle\langle 1|, \nonumber \\
|0\rangle\langle 1| & \longrightarrow & \frac{R_2}{4}         |0\rangle\langle 0| +  \frac{2 + R_1 - 4W}{4}|0\rangle\langle 1|
                                      - \frac{2 - R_1 - 4W}{4}|1\rangle\langle 0| - \frac{R_2}{4}         |1\rangle\langle 1|, \nonumber \\
|1\rangle\langle 0| & \longrightarrow & \frac{R_2}{4}         |0\rangle\langle 0| - \frac{2 - R_1 - 4W}{4}|0\rangle\langle 1|
                                      + \frac{2 + R_2 - 4W}{4}|1\rangle\langle 0| - \frac{R_2}{4}         |1\rangle\langle 1|, \nonumber \\
|1\rangle\langle 1| & \longrightarrow & \frac{2 - R_1}{4}|0\rangle\langle 0| + \frac{R_2}{4}    |0\rangle\langle 1| 
                                      + \frac{R_2}{4}    |1\rangle\langle 0| + \frac{2 + R_1}{4}|1\rangle\langle 1|.
\end{eqnarray}
The gate fidelity averaged over all possible $|\psi\rangle$ is given by
\begin{eqnarray}\label{sigmayfidelity}
{\cal G}_{\alpha}(\frac{\pi}{2}) & \equiv & \max_{t = \tau_{\alpha}(\pi/2)}\left\{\frac{1}{4\pi}\int^{2\pi}_0\int^{\pi}_0
\langle\phi_{\alpha}|{\cal E}^{\alpha}(|\psi\rangle\langle\psi|)|\phi_{\alpha}\rangle\sin\theta d\theta d\phi\right\} \nonumber \\
& = & \max_{t = \tau_{\alpha}(\pi/2)}\left\{\frac{1}{6}(4 - R_2 - 2W)\right\}.
\end{eqnarray}
Here, $|\phi_{\alpha}\rangle = {\cal R}^{(\hat{\alpha})}(\pi/2)|\psi\rangle$ and $\alpha = x$ or $y$.
\begin{table}
\begin{tabular}{|c|c|c|c|c|c|c|} \hline
$B_0$/mT                                           & 1    & 2    & 3    & 4    & 5    & 6    \\ \hline
${\cal G}_{\alpha}(\pi/2)$                         & 0.88 & 0.96 & 0.98 & 0.99 & 0.99 & 0.99 \\ \hline
Duration, $\tau_{\alpha}(\pi/2)$/ns                & 6.1  & 4.0  & 2.8  & 2.2  & 1.8  & 1.5  \\ \hline
\end{tabular} \\
Table IV: \quad Gate fidelities for different $B_0$'s. \par
\end{table}
Comparing Table IV with Table I, we observe that $\tau_{\alpha}(\pi) \approx 2\tau_{\alpha}(\pi/2)$ for $B_0 \geq 2$.  For instance, with 
$B_0 = 6 {\rm mT}$, it takes about 1.5 ns to reach a gate fidelity of 0.99.  However, when $B_0 = 1\ {\rm mT}$, we note a deviation from the above 
observation.  It is a manifestation of the nonlinearity of open-system dynamics, which becomes apparent if $B_0$ is small and the respective 
$\tau_{\alpha}(\pi/2)$ and $\tau_{\alpha}(\pi)$ are large.  This nonlinearity also affects the composition of gates.  
To see this, we calculate the composite gate fidelity
\begin{eqnarray}\label{rotfidelity}
{\cal U}_{jk} & \equiv & \max_{t = \tau_{\rm com}}\left\{\frac{1}{4\pi}\int^{2\pi}_0\int^{\pi}_0
\langle\psi|U^{\dagger}_{jk}\tau^{jk}U_{jk}|\psi\rangle\sin\theta d\theta d\phi\right\} \nonumber \\
& = & \max_{t = \tau_{\rm com}}\left\{\frac{1}{2} + \frac{1}{8}\left[R^2_2(1 - 2W) \pm \frac{1}{6}R^3_1\right]\right\}.
\end{eqnarray}
In Eq.(\ref{rotfidelity}), $+$ applies when $jk = 00$ or $11$, and $-$ when $jk = 01$ or $10$.  And,
\begin{eqnarray}
\tau^{00} & = & {\cal E}^x_+({\cal E}^y_+({\cal E}^x_-(|\psi\rangle\langle\psi|))), \nonumber \\
\tau^{01} & = & {\cal E}^x_+({\cal E}^y_+({\cal E}^x_+(|\psi\rangle\langle\psi|))), \nonumber \\
\tau^{10} & = & {\cal E}^x_+({\cal E}^y_-({\cal E}^x_+(|\psi\rangle\langle\psi|))), \nonumber \\
\tau^{11} & = & {\cal E}^x_-({\cal E}^y_+({\cal E}^x_+(|\psi\rangle\langle\psi|))). \nonumber
\end{eqnarray}
${\cal E}^{\alpha}_+ = {\cal E}^{\alpha}$, and ${\cal E}^{\alpha}_- = {\cal E}^{\alpha}$ with $R_2 \rightarrow -R_2$ when $B_0 \rightarrow -B_0$.  
Clearly, ${\cal U}_{jk} \not= [{\cal G}_{\alpha}(\pi/2)]^3$.  This is obviously true when $B_0 = 1\ {\rm mT}$, where in fact ${\cal U}_{jk} > 
[{\cal G}_{\alpha}(\pi/2)]^3$.  However, for large enough $B_0$ ($\geq 2\ {\rm mT}$), we have ${\cal U}_{jk} \approx [{\cal G}_{\alpha}(\pi/2)]^3$.

\begin{table}
\begin{tabular}{|c|c|c|c|c|c|c|} \hline
$B_0$/mT                                                                & 1           & 2    & 3    & 4    & 5    & 6    \\ \hline
${\cal U}_{00}$ or ${\cal U}_{11}$ (${\cal U}_{01}$ or ${\cal U}_{10}$) & 0.73 (0.71) & 0.88 & 0.94 & 0.96 & 0.98 & 0.98 \\ \hline
Duration, $\tau_{\rm com}$/ns                                           & 16 (19)     & 12   & 8.5  & 6.5  & 5.4  & 4.4  \\ \hline
\end{tabular} \\
Table V: \quad Composite gate fidelities for different $B_0$'s. \par
\end{table}

Finally, to determine the impact of the nuclear spin baths on the protocoal $\cal T$, we calculate the teleportation fidelity
\begin{equation}
\Phi \equiv \max_{t = \tau_{\rm tot}}\left\{\frac{1}{4\pi}\int^{\pi}_0\int^{2\pi}_0
\sum^1_{i, j = 0}p_{ij}{_B}\langle\psi|\rho^{ij}_B|\psi\rangle_B\sin\theta d\theta d\phi\right\},
\end{equation}
where
\begin{eqnarray}
\rho^{00} & = & {\cal E}^x_+({\cal E}^y_+({\cal E}^x_-(\sigma^{00}(\tau)))), \nonumber \\
\rho^{01} & = & {\cal E}^x_+({\cal E}^y_+({\cal E}^x_+(\sigma^{01}(\tau)))), \nonumber \\
\rho^{10} & = & {\cal E}^x_+({\cal E}^y_-({\cal E}^x_+(\sigma^{10}(\tau)))), \nonumber \\
\rho^{11} & = & {\cal E}^x_-({\cal E}^y_+({\cal E}^x_+(\sigma^{11}(\tau)))).
\end{eqnarray}
From Table VI, we conclude that one can achieve nonclassical teleportation fidelities of greater than $2/3$ \cite{Horodecki, Bose} when 
$B_0 \geq 2$ mT.  Together with Table III and Table V, we deduce that $\Phi \approx {\cal F} \times [{\cal G}_{\alpha}(\pi/2)]^3$ and 
$\tau_{\rm tot} \approx \tau + \tau_{\rm com}$, with the approximations becoming better with increasing $B_0$.

\begin{table}
\begin{tabular}{|c|c|c|c|c|c|c|} \hline
$B_0$/mT                             & 1    & 2     & 3     & 4     & 5    & 6    \\ \hline
$\Phi$                               & 0.53 & 0.67  & 0.79  & 0.86  & 0.90 & 0.93 \\ \hline
Total Duration, $\tau_{\rm tot}$/ns  & 31   & 20.2  & 14.2  & 10.9  & 8.8  & 7.3  \\ \hline
\end{tabular} \\
Table VI: \quad Teleportation fidelities for different $B_0$'s. \par
\end{table}

In conclusion, we have carried out the first detailed study of the decohering effects of nuclear spin baths on the teleportation protocol $\cal T$, 
proposed by de Visser and Blaauboer in Ref.\cite{Visser}.  We give the completely positive, trace-preserving maps that describe the open-system dynamics
of electron spins in a semiconductor nanostructure consisting of a single and a double quantum dot.  This allows us to identify the single-qubit 
measurements as the source of noise amplification.  It also enables us to pinpoint when nonlinearities of the quantum evolutions become significant.  
These are important points to note when one adapts our studies to the practical situation where $B_0$ is not in the form of a square pulse.  They must 
also be considered in feasibility studies of quantum information processing proposals involving quantum dots, such as that in Ref.\cite{Blaauboer}.  Our 
results show that it is possible to achieve nonclassical teleportation fidelity with $\cal T$ in the presence of decoherence caused by the nuclear spin 
baths.  The technical challenge lies in the creation of an ESR square pulse of arbitrary amplitude.


\begin{thebibliography}{99}
\bibitem{Bennett} C. H. Bennett, G. Brassard, C. Crepeau, R. Jozsa, A. Peres, and W. K. Wootters, Phys. Rev. Lett. {\bf 70}, 1895 (1993).
\bibitem{Gottesman} D. Gottesman and I. L. Chuang, Nature {\bf 402}, 390 (1999).
\bibitem{Visser} R. L. de Visser and M. Blaauboer, Phys. Rev. Lett. {\bf 96}, 246801 (2006).
\bibitem{Nielsen} M. A. Nielsen and I. L. Chuang, {\em Quantum Computation and Quantum Information} (Cambridge University Press, Cambridge, England, 2000).
\bibitem{DiVincenzo} D. Loss and D. P. DiVincenzo, Phys. Rev. A {\bf 57}, 120 (1998).
\bibitem{Johnson} A. C. Johnson, {\it et al.}, Nature (London) {\bf 435}, 925 (2005).
\bibitem{Koppens} F. H. L. Koppens, {\it et al.}, Science {\bf 309}, 1346 (2005).
\bibitem{Petta} J. R. Petta, {\it et al.}, Science {\bf 309}, 2180 (2005).
\bibitem{Wenxian} W. Zhang, V. V. Dobrovitski, K. A. Al-Hassanieh, E. Dagotto, and B. N. Harmon, Phys. Rev. B {\bf 74}, 205313 (2006).
\bibitem{Horodecki} M. Horodecki, P. Horodecki and R. Horodecki, Phys. Rev. A {\bf 60}, 1888 (1999).
\bibitem{Bose} S. Bose and V. Vedral, Phys. Rev. A {\bf 61}, 040101(R) (2000).
\bibitem{Blaauboer} F. Bodoky and M. Blaauboer, Phys. Rev. A {\bf 76}, 052309 (2007).
\end{thebibliography}
\end{document}